\documentclass[aps,prl,preprint,superscriptaddress]{revtex4}

\usepackage{amsmath,bm}
 \usepackage{mathrsfs}
 \usepackage{amsfonts}
 \usepackage{graphicx}

  \usepackage{ulem}
 \usepackage{setspace} %leaves captions single space in draft mode
 \usepackage{epstopdf}
 \usepackage{dcolumn}
 \usepackage{amsmath}
 \usepackage{epsfig}
 \usepackage{indentfirst}
 \usepackage{psfrag}
 \usepackage{subfigure}
 \usepackage{float}
 \usepackage{amssymb}
 \usepackage{color}
 \usepackage{units} % rz
 \usepackage{graphicx}% Include figure files
 \usepackage{dcolumn}% Align table columns on decimal point
 \usepackage{bm}% bold math
 \usepackage{physics}
 \usepackage{dcolumn}% Align table columns on decimal point
 \usepackage{bm}% bold math
 \usepackage{natbib}
 \usepackage{xr-hyper}
\usepackage[backref=none,bookmarksnumbered=true,bookmarks=true,bookmarksopen=true,colorlinks=true,citecolor=blue,linkcolor=blue,filecolor=blue,anchorcolor=green,urlcolor=blue,unicode=false]{hyperref}

  \makeatletter
  \newcommand*{\addFileDependency}[1]{% argument=file name and extension
  \typeout{(#1)}
  \@addtofilelist{#1}
  \IfFileExists{#1}{}{\typeout{No file #1.}}
}
\makeatother
\newcommand*{\myexternaldocument}[1]{%
    \externaldocument{#1}%
    \addFileDependency{#1.tex}%
    \addFileDependency{#1.aux}%
}
 \myexternaldocument{ChiralTopo-SI}
\newcommand{\rev}[1]{{\color{black} #1}}

\makeatletter
\newcommand\colorsout[1]{\bgroup \markoverwith{\textcolor{#1}{\rule[0.5ex]{2pt}{0.4pt}}}\ULon}

\makeatother

\makeatletter
\let\saved@includegraphics\includegraphics
\AtBeginDocument{\let\includegraphics\saved@includegraphics}
\renewenvironment*{figure}{\@float{figure}}{\end@float}
\makeatother
\begin{document}
\title{Topological phase transition in chiral graphene nanoribbons: \\from edge bands to end states}
   
\setstretch{1.2} 
\author{Jingcheng Li$^*$}
    \affiliation{CIC nanoGUNE-BRTA, 20018 Donostia-San Sebasti\'an, Spain}
\author{Sofia Sanz$^*$}
	\affiliation{Donostia International Physics Center (DIPC), 20018 Donostia-San Sebasti\'an, Spain}
\author{Nestor Merino-D\'{i}ez$^{*}$}
    \affiliation{CIC nanoGUNE-BRTA, 20018 Donostia-San Sebasti\'an, Spain}
	\affiliation{Donostia International Physics Center (DIPC), 20018 Donostia-San Sebasti\'an, Spain}
    \affiliation{Centro de F\'{\i}sica de Materiales CSIC-UPV/EHU, 20018 Donostia-San Sebasti\'an, Spain}
\author{Manuel Vilas-Varela$^*$}
    \affiliation{Centro Singular de Investigaci\'on en Qu\'imica Biol\'oxica e Materiais Moleculares (CiQUS), and Departamento de Qu\'imica     Org\'anica, Universidade de Santiago de Compostela, Spain}
\author{Aran Garcia-Lekue}
	\affiliation{Donostia International Physics Center (DIPC), 20018 Donostia-San Sebasti\'an, Spain}
    \affiliation{Ikerbasque, Basque Foundation for Science, 48013 Bilbao, Spain}
\author{Martina Corso}
    \affiliation{Centro de F\'{\i}sica de Materiales CSIC-UPV/EHU, 20018 Donostia-San Sebasti\'an, Spain}
\author{Dimas G. de Oteyza $^{\bigtriangleup}$}
	\affiliation{Donostia International Physics Center (DIPC), 20018 Donostia-San Sebasti\'an, Spain}
    \affiliation{Ikerbasque, Basque Foundation for Science, 48013 Bilbao, Spain}
\author{Thomas Frederiksen$^{\bigtriangleup}$}
	\affiliation{Donostia International Physics Center (DIPC), 20018 Donostia-San Sebasti\'an, Spain}
    \affiliation{Ikerbasque, Basque Foundation for Science, 48013 Bilbao, Spain}
\author{Diego Pe{\~{n}}a$^{\bigtriangleup}$}
    \affiliation{Centro Singular de Investigaci\'on en Qu\'imica Biol\'oxica e Materiais Moleculares (CiQUS), and Departamento de Qu\'imica Org\'anica, Universidade de Santiago de Compostela, Spain}
\author{Jose Ignacio Pascual$^{\bigtriangleup}$}
    \affiliation{CIC nanoGUNE-BRTA, 20018 Donostia-San Sebasti\'an, Spain}
    \affiliation{Ikerbasque, Basque Foundation for Science, 48013 Bilbao, Spain}
\vspace{1cm}

\begin{abstract}
\setstretch{1.2} 
Precise control over the size and shape of graphene nanostructures allows engineering spin-polarized edge and topological states, representing a novel source of non-conventional $\pi$-magnetism with promising applications in quantum spintronics. A prerequisite for their emergence is the existence of robust gapped phases, which are difficult to find in extended graphene systems:   only armchair graphene nanoribbons (GNRs) show a band gap that, however, closes for any other GNR orientation. Here we show that semi-metallic chiral GNRs (chGNRs) narrowed down to nanometer widths undergoes a topological phase transition, becoming first topological insulators, and transforming then into trivial band insulators for the narrowest chGNRs. We fabricated atomically precise chGNRs of different chirality and size by on surface synthesis using predesigned molecular precursors. Combining scanning tunnelling microscopy (STM) measurements and theory simulations, we follow the evolution of topological properties and bulk band gap depending on the width, length, and chirality of chGNRs. The first emerging gapped phases are topological, protected by a chiral interaction pattern between edges. For narrower ribbons, the symmetry of the interaction pattern changes, and the topological gap closes and re-opens again as a trivial band insulator. Our findings represent a new platform for producing topologically protected spin states and demonstrates the potential of connecting chiral edge and defect structure with band engineering.

$*$ These authors contributed equally to this work. 
  \end{abstract}
   
\maketitle

\setstretch{1.2} % Linespacing text
%\baselineskip19pt % Linespacing text
%\linespread{1.7} % Linespacing Captions

\newpage

Band topological classification of materials has been successfully applied to predict and explain the emergence of exotic states of matter such as Quantum Spin Hall (QSH) edge states in topological insulators~\cite{Hasan2010,Qi2011,Kane2005} or topological superconductivity \cite{Sato2017}. The potential of this classification relies on the protection of the topological order by a symmetry, that can undergo a topological phase transition when the symmetry is changed. Symmetry Protected Topological (SPT) phase transitions were observed in artificial semiconducting systems such as two-dimensional quantum wells of HgTe\cite{Konig2007} and one-dimensional organic polymers \cite{Cirera2020}. The key element is the existence of two gapped SPT phases, the traditional (trivial) band insulator and the non-trivial  topological insulating state,   separated by a metallic state.  

In spite of being a semimetal, graphene has the potential to build up SPT phases by opening an energy  gap around the Fermi level and endowing the lattice with an additional chiral symmetric interaction \cite{Haldane1988}. For example,  one-dimensional SPT phases were engineered inside the band-gap of armchair GNRs by modelling a one-dimensional Su-Schrieffer-Heeger (SSH) chain \cite{SSH1979}  with edge moieties containing localized in-gap states \cite{Cao2017,Groning2018,Rizzo2018,Ruffieux2020,Rizzo2020}. In zigzag GNRs (ZGNRs), however, the existence of zero-energy edge bands  \cite{Nakada1996,Fujita1996} brings the system into a metallic state that  prevents the appearance of gaped topological phases.  As proposed by Kane and Mele \cite{Kane2005},  the presence of spin-orbit interaction can open a gap in bulk graphene and turn the zero-energy  modes into QSH edge states. However, the expected gap induced by spin orbit interaction in graphene is very small, and this effect could only be present at very low temperatures \cite{Blick2019}.

Here, we demonstrate that sizeable topological insulating phases emerge in narrow chiral GNRs driven by the interaction between the opposing edges. The term chiral GNRs (chGNRs) refers to the large set of ribbons extending along low-symmetry crystallographic directions ($n,m$) of graphene. The genuine zero-energy  edge bands of ZGNRs persist in chGNRs via the accumulation of states at zero energy over zigzag sites, including their spin polarization in the presence of Coulomb electron-electron interactions~\cite{Yazyev2011,Tao2011}. However, chiral ribbons are more sensitive to a reduced width than ZGNRs, allowing to easily produce gapped GNRs with inherited chirality from the edge reconstruction.  

%*****************************FIGURE 1***********************************

\begin{figure}[!t]
\centering
\includegraphics[width=0.99\columnwidth]{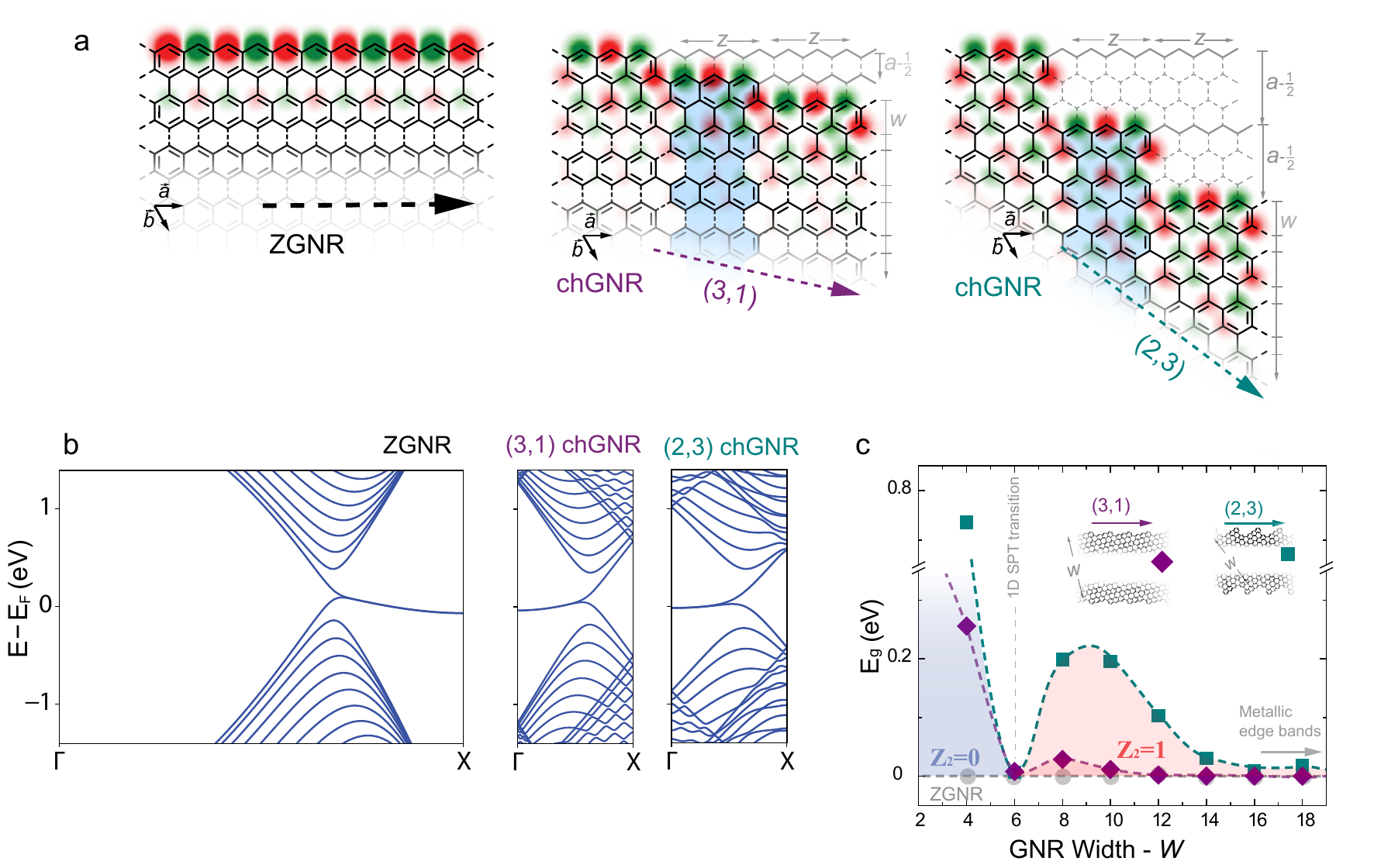}
\caption{\textbf{Edge states in zigzag and chiral GNRs:}. 
\textbf{a}  Edge structure of wide zigzag GNRs,  and chiral GNRs with chiral vectors (3,1) and (2,3). Superimposed,  tight-binding (TB) wave-function amplitude and spin polarization of their zero energy modes. The  blue-shaded area indicates the rectangular graphene building block used for constructing a family of chiral GNRs, with edge structure alternating  \textit{z} zigzag and $a$-1/2 armchair segments. For the both chiral ribbons $z=3$, while $a=1$ and 2, respectively  (details in the text and in Supplementary Note \ref{SN:STRUCT}).
\textbf{b}  Band structure of 40-carbon wide ZGNR, and (3,1) and (2,3) chiral GNRs from TB simulations. The slight dispersion of the flat bands is due to the inclusion of second-nearest neighbour interactions in our model \ref{SN:TBWF}.
\textbf{c} Band gap values of (3,1) and (2,3) chiral GNRs (purple diamond and blue squared symbols), as a function of the width (grey circles indicate the zero gap of ZGNRs). The value of the computed topological invariant  $\mathbb{Z}_2$ is indicated. A  topological transition occurs for ribbons at the width $w=6$.
} \label{fig0}
\end{figure}
%********************************************************

\textbf{Prediction of a SPT phase transition:} 
We consider  a family of chGNRs customized from a basic rectangular aromatic block of length \textit{z} (number of zigzag unit cells) and width $w$ (number of carbon atoms across), blue-shadowed in Fig.~\ref{fig0}a. Chiral GNRs along any chiral direction ($n,m$) can be obtained simply by repeating and shifting these blocks along their armchair edges an amount of $a-1/2$ armchair unit cells, and connecting them with C-C bonds. The edges of the resulting chGNR alternate \textit{z} zigzag  and \textit{a}-1/2  armchair sites to accommodate its orientation to the chiral vector, such that $(n,m)=(z+1-a,2a-1)$ (see Supplementary Note~\ref{SN:STRUCT}). This edge reconstruction promotes the localization of zero-energy edge states at the $z$ zigzag segments, while the perpendicular armchair spacers act as potential barriers between them \cite{Santos2013}. 
For example, our tight-binding (TB) simulations  in Fig.~\ref{fig0}a,b for wide chGNRs (theory methods in Supplementary Part \ref{SN:TBwide}) show the presence of zero-energy bands localized at the zigzag edge segments, reminiscent of the edge bands in the zigzag edges of graphene, which decay towards the center of the ribbon. Therefore, these ribbons lie in a metallic phase, with electron mobility that can be described as hopping between zigzag segments along the chGNR edge.

However, this metallic phase vanishes in narrow ribbons due to hybridization of bands at opposing edges. Interestingly, the emergent gapped phase  corresponds to a SPT insulating phase characterized by an invariant $\mathbb{Z}_2=1$, as obtained from the computed Zak phase $\gamma_z=\pi$ \cite{Zak1989} of the occupied bands (using a rectangular unit cell enclosing the blue block in Fig.~\ref{fig0}a). The non-trivial topological class of this phase turns out to be a global property of narrow chGNRs of this family, protected by the symmetry of a chiral hybridization pattern between edges. 

Our simulations also find that the topological phase vanishes upon further reducing the width of the ribbon. The gap closes and reopens again as a trivial band insulator, characterized by $\mathbb{Z}_2=0$. In agreement with the properties of SPT phases, this new trivial state is connected with a symmetry change in the interaction pattern between the edges. 
Figure~\ref{fig0}c exemplifies the transition for ribbons with ($n=3$,$m=1$) and (2,3) chiral vectors, by plotting their evolution of the energy gap with the width $w$. First, a sizable energy gap opens up as the ribbons are narrowed down, which corresponds to a one-dimensional topological insulating phase, and at a critical width of $w\sim6$, the gap closes, and reopens, now in a trivial phase.   

%************************************FIGURE 1****************************
\begin{figure}[!t]
\centering
\includegraphics[width=\columnwidth]{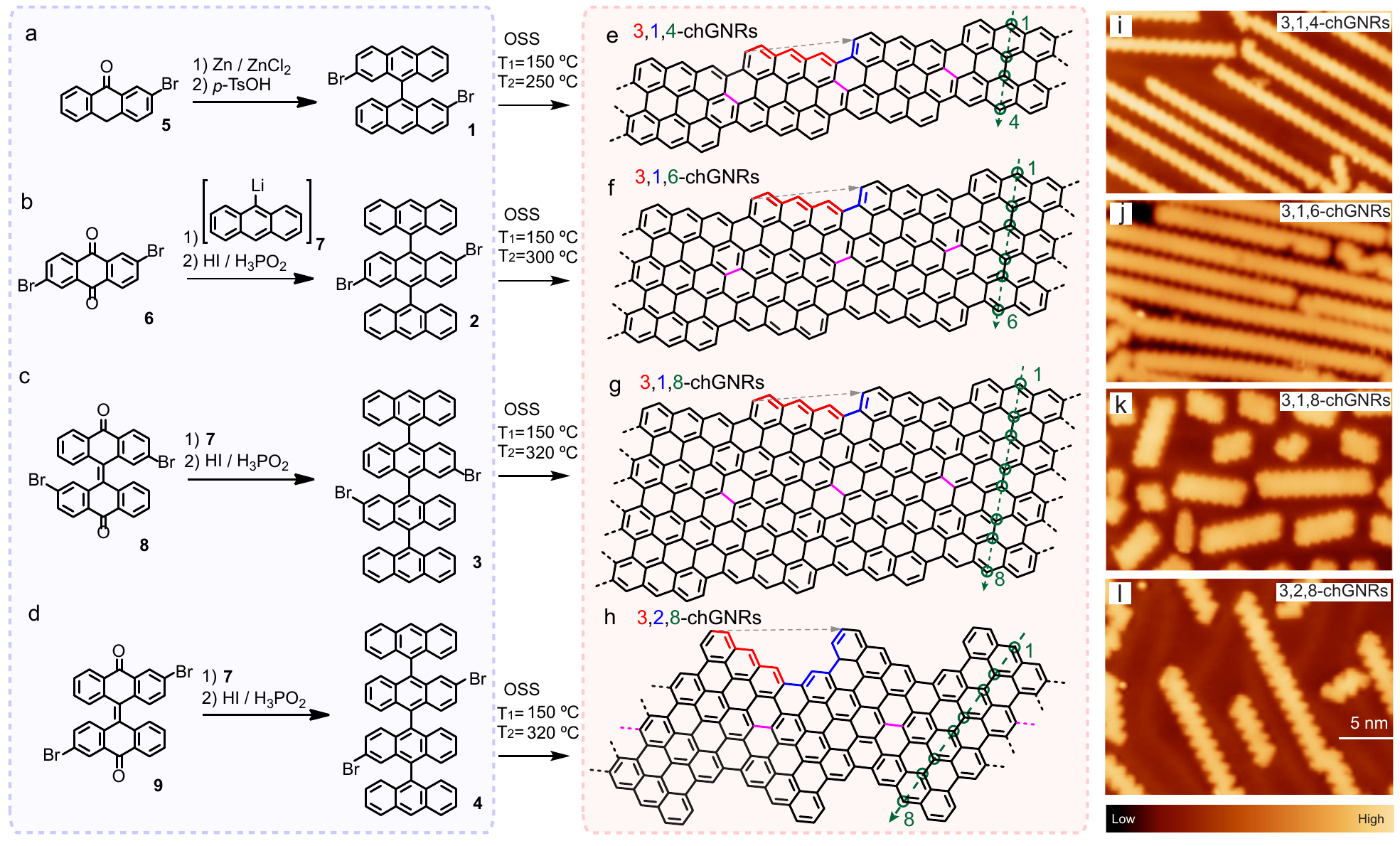}
\caption{\textbf{Synthetic strategy to produce chGNRs \rev{combining solution and on-surface synthesis}}. \textbf{a-d}, solution synthesis protocols for producing molecular precursors \textbf{1, 2, 3}, for the synthesis of 3,1,$w$-chGNRs  with different widths ($w=4$, 6 and 8), and  precursor \textbf{4} for 3,2,8-chGNRs. Synthesis details are described in the text and in the Methods section.
\textbf{e-h}, chemical structures of chGNRs after  Ullmann coupling and CDH of the four molecular precursors in \textbf{a-d}, respectively. Red(blue) carbon bonds highlight the zigzag(armchair) segments of the chGNRs. Green circles highlight the carbon atoms across the GNRs, which measure the width of the GNRs. The GNRs are  named following the sequence of $z$ zigzag sites, \textit{a} brominated site (see \ref{SN:STRUCT}), and $w$ width, as $z,a,w$-chGNRs.  \textbf{i-l}, STM overview images of the formed chGNRs on a Au(111) surface after Ullmann coupling and CDH of the corresponding molecular precursors \textbf{1-4} (sample bias $V_s=1$ V, scale bar as labeled in \textbf{l}).
}   \label{OSS}
\end{figure}
%********************************************************

\textbf{Fabrication of chGNRs:} To demonstrate the predicted topological phase transition described above, we fabricated several members of this chGNR family with different chiral vectors and widths (Fig.~\ref{OSS}) though a combination of customized organic precursors and on-surface synthesis over a gold (111) surface~\cite{Talirz2016,Clair2019}.  
Our strategy started with the synthesized poli-[n']anthracene precursor molecules \textbf{1}, \textbf{2}, \textbf{3}, and \textbf{4} shown in Fig.~\ref{OSS}a-d for the synthesis of chGNRs with $z=3$ edge structure. Their customized structures composed of an increasing number of anthacene units and Br functionalization sites \textit{a}, were designed for obtaining (3,1) chGNRs with different widths (using \textbf{1, 2, 3}) and chiral angle (e.g. (2,3) chGNR, using \textbf{4}) through a sequence of OSS steps. For a fixed $z$, the  Br-substitution site, labelled \textit{a} in \ref{SN:STRUCT}, determines the ($n,m$) chiral vector by  steering the  Ullmann-like connection between molecular precursors with a shift of $a-1/2$ armchair unit lengths (see Fig.~1a). The number of anthracene units [n'] determines the  width $w=2$[n'] of the ribbon. Hence, in the following we label  the ribbons of this family as \textit{z,a,w}-chGNR (\ref{SN:STRUCT}).

The GNR precursors were prepared in solution, as shown in Fig.~\ref{OSS}a-d. Compound \textbf{1}, which is formed by the linking of two anthracenes and constitutes the molecular precursor of 3,1,4-chGNR, was obtained by Zn-promoted reductive coupling of bromoanthrone \textbf{5}, followed by dehydration~\cite{DeOteyza2016}. The trisanthracene \textbf{2}, precursor of 3,1,6-chGNRs, was obtained in one pot from dibromoanthraquinone \textbf{6}, by addition of two equivalents of 9-anthracenyl lithium (\textbf{7}) followed by reduction with a mixture of HI and H$_3$PO$_2$. Similarly, the tetrakisanthracene \textbf{3}, precursor of 3,1,8-chGNRs, was synthesized by reaction of compound \textbf{8} with organolithium \textbf{7}, followed by a reduction step. Finally, compound \textbf{4}, precursor of 3,2,8-chGNRs, was prepared from derivative \textbf{9} following a similar procedure (see Supplementary Part~\ref{SN:synthesis} for further details).
%The precursors were prepared in solution, as schematized in Fig.~\ref{OSS}. The key steps are based in the reductive coupling of different bromoanthrones cores (5-9) obtained with the target halogenation site A. For  precursor \textbf{1}, the coupling of cores 5 is promoted by Zn \cite{DeOteyza2016}. For the rest, two 9-anthracenyl lithium molecules \textbf{7} are coupled to the cores, reducing the resulting diol in HI/H3PO2 (see Methods and SN2 for further details on the solution synthesis). 
In the OSS step, each precursor was independently sublimated onto a clean Au(111) surface held at room temperature and step-wisely annealed to $T_1$ to induce their Ullmann-like polymerization. Subsequently, a further annealing to $T_2$ activated the cyclodehydrogenation (CDH) of the polymers into the targeted chiral graphene nanoribbons with chiral vectors (3,1) and (2,3) and with different widths (Fig.~\ref{OSS}(e-h)).
STM images of the resulting structures (Fig.~\ref{OSS}i-l show the characteristic straight, and planar shape of the ribbons, confirming their successful synthesis.  

%******************************** FIGURE 3*****************
\begin{figure}[t]
\centering
\includegraphics[width=0.99\textwidth]{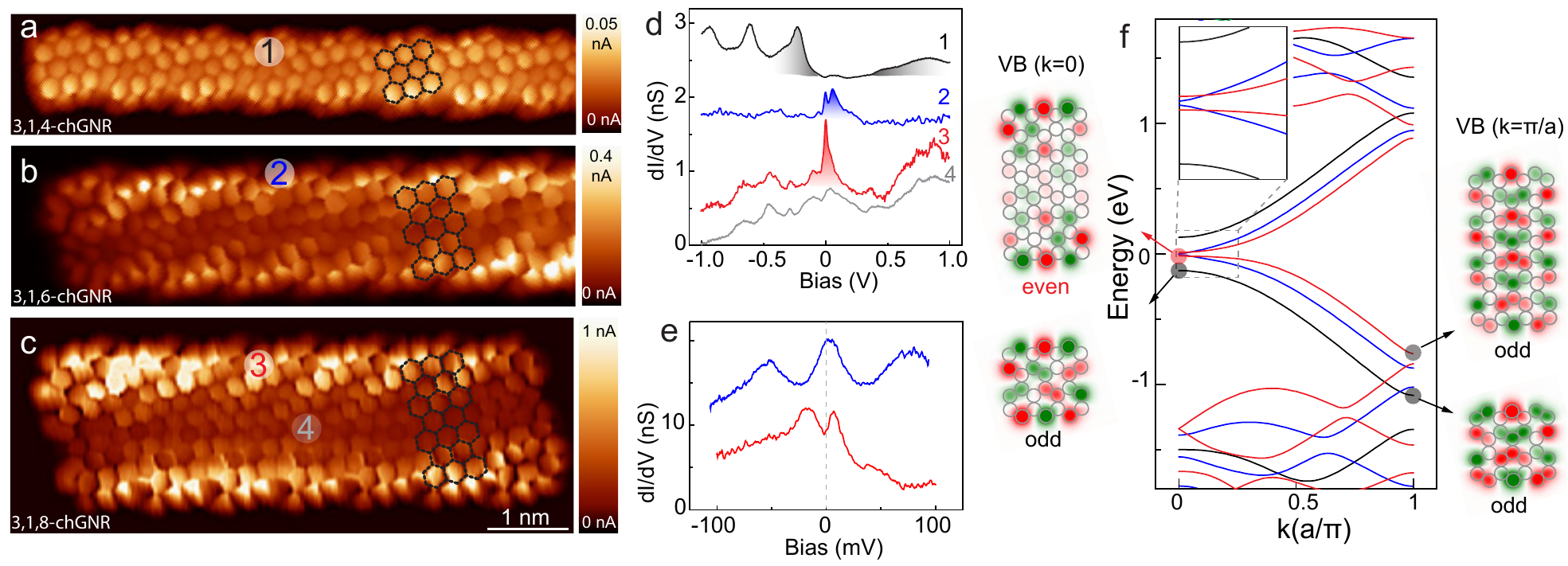}
\caption{\textbf{Emergence of edge bands and gaps in 3,1,\textit{w}-chGNRs}. \textbf{a-c}, Bond-resolved constant height $dI/dV$ images of 3,1,4-, 3,1,6-, and 3,1,8-chGNRs, respectively ($V= 2$ mV). The images were acquired with a CO-terminated tip. All the images share the same scale bar as labeled in c. The  chemical structure of the basic aromatic block is superimposed on the images.  \textbf{d,e}, $dI/dV$ spectra taken at the locations as noted on the images. The spectra are shifted for clarity. \textbf{f}, Band diagrams in black, blue and red are the TB simulated band structure of 3,1,4-, 3,1,6- and 3,1,8-chGNRs, respectively. Inset shows a zoom of the bands around zero energy. The wavefunction at $\Gamma$ ($k=0$) and X ($k=\pi/a$) of VB of 3,1,4, and 3,1,8-chGNRs are shown in the left and right panels in the figure, with their inversion symmetry indicated.}  \label{sts}
\end{figure}
%*************************************************************

\textbf{Emergence of edge bands in wide 3,1,\textit{w}-chGNRs:} 
We compare first the effect of increasing the width on the electronic structure of 3,1,\textit{w}-chGNRs. Bond-resolved STM images shown in Fig.~\ref{sts}a-c (obtained by measuring constant height current maps at $V=2$ mV using a CO-terminated tip \cite{Gross2009,Kichin2011}) reproduce the hexagonal ring patterns of the different ribbons, in agreement  with the  chemical structures in Fig.~\ref{OSS}g-i. However,  the wider 3,1,6- and 3,1,8-chGNRs show, on top of the ring structure, a characteristic current increase over the edges, which is absent in the 3,1,4-chGNR images. 
These brighter edges unveil a larger density of states (DOS) around the Fermi energy, this being an experimental evidence for the emergence of edge bands in the wider ribbons. This is further corroborated by comparing differential conductance spectra ($dI/dV$) on the different ribbons, as shown in Fig.~\ref{sts}d. 
The spectral plots 2 and 3, measured at the edges of 3,1,6- and 3,1,8-chGNRs, respectively, show a pronounced increase of $dI/dV$ signal around zero bias, that is absent over the central part of the ribbons (spectral plot 4). 
In contrast, a wide band gap of $\sim$ 0.7 eV with no DOS enhancement around the Fermi level is found all over the 3,1,4-chGNRs (plot 1 in Fig.~3d) \cite{Merino-Diez2018}.

The emergence of edge bands close to zero-energy in 3,1,6- and 3,1,8-chGNRs is reproduced by our TB  (Fig.~\ref{sts}f) and  density functional theory simulations (\ref{SN:DFTband}) of the band structure of infinitely long 3,1,$w$-chGNRs. The relatively large band gap ($E_g$=0.26 eV) of the 3,1,4-chGNR closes abruptly for the  wider ribbons, whose valence and conduction bands (VB and CB) apparently merge at zero energy and flatten, being these the edge-localized states resolved in the experiments. 
However, as we show in the inset of Fig.~\ref{sts}f, the frontier bands of 3,1,6- and 3,1,8-chGNRs do not overlap at zero, but remain gapped. Contrary to a monotonous gap closing, the theoretical energy gap is very small for 3,1,6-chGNRs ($\sim$8 meV), and opens again for the wider 3,1,8-chGNRs ($\sim$29~meV). Only for $w\geq12$ the gap closes definitively (see Fig.~\ref{fig0}c). 

The origin of the mini-gap reopening from $w=6$ to $w=8$ is connected with a gap inversion due to a change in valence band's topology with the width. This can be deduced by comparing maps of the wave function amplitude and phase distribution at $k=0$ and at $k=\pi/a$, shown in Fig.~\ref{sts}f.  
The VB of 3,1,4-chGNRs (\ref{SN:TBWF}) maintains an odd inversion symmetry with respect to the center of the unit cell in all $k$-space, whereas for 3,1,8-chGNRs it changes parity (from even, at $k=0$, to odd at $k=\pi/a$), revealing a band inversion at $k=0$. As a consequence, the wave function acquires a net phase as it disperses along the Brillouin space. To connect these differences in parity with topological classes,  we computed the Zak phase $\gamma_z$ \cite{Zak1989} of the occupied band structure for every ribbon  and obtained their $\mathbb{Z}_2$ invariant,  as described in \ref{SN:ZAK}. The 3,1,4-chGNRs accumulate a global Zak phase of $\gamma_z=0$ and, hence, are in a trivial topological phase, i.e. $\mathbb{Z}_2$=0. The intermediate case, 3,1,6-chGNRs, has a very small gap that changes topological phase depending on details of the simulation (\ref{SN:TBWF}) and thus, we consider here as the transition  metallic case. 
For the wider ribbon, 3,1,8-chGNR, we obtain $\gamma_z=\pi$ in accordance with a  non-trivial SPT phase ($\mathbb{Z}_2=1$), thus accounting for the gap reopening found in the simulations.

 %*****************************FIGURE 4 *****************************
 
\begin{figure}[tb!]
\centering
\includegraphics[width=0.9\textwidth]{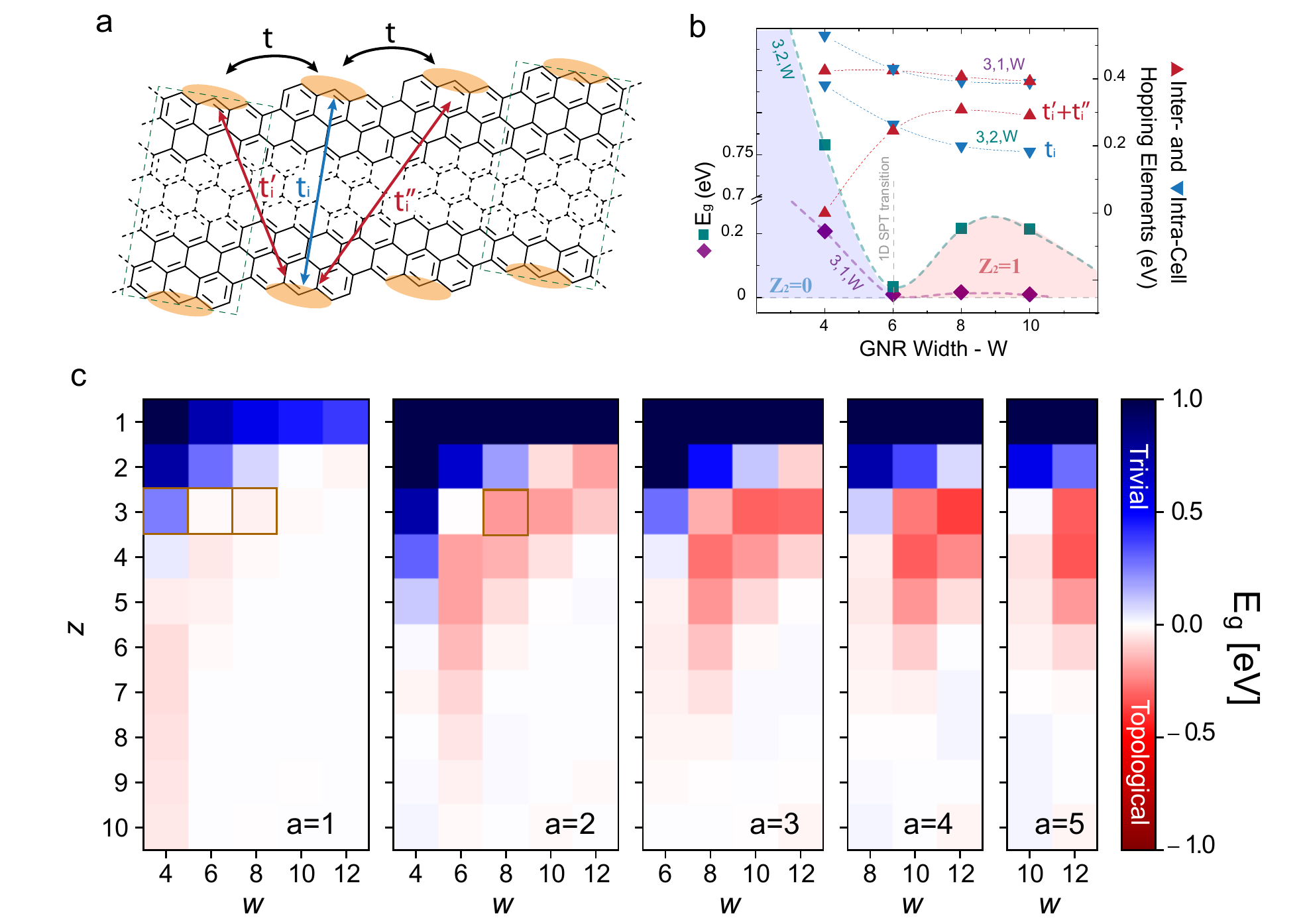}
\caption{\textbf{SSH model and simulations of the band topology of chGNRs: } \textbf{a}, Modified Su-Shrieffer-Heeger model to explain the SPT phase transition of chGNRs with the width. 
\textbf{b}, Evolution of the energy gap value and sign with the chGNR width, obtained by fitting the mSSH model  in \textbf{a} to frontier bands from 3NN TB simulations. (Left ordinate) comparison of intra- ($t_{i}^{0}$) and inter-edge ($t_{i}^{-1}+t_{i}^{+1}$) hopping elements obtained from the fit, indicating that the topological transition occurs when both are balanced. \textbf{c} Color plots of the energy gap value and sign for the family of $(z,a,w)$ chGNRs, showing that the topological transition is a global property of this family of chiral ribbons. Marked squares indicate the ribbons studied in this work.   The red/blue color scale refers to the positive/negative sign of $E_g$, defined as $(-1)^{\mathbb{Z}_2}$, where the $\mathbb{Z}_2$ invariant is obtained from the Zak phase (\ref{SN:ZAK}).   
}  \label{ssh}
\end{figure}
 %*****************************  *****************************
 
\textbf{Su-Shrieffer-Heeger model prediction of a SPT phase transition:} The presence of band gaps in narrow chGNRs and the SPT phase transition can be explained using the modified Su-Shrieffer-Heeger (mSSH) model~\cite{SSH1979} depicted in Figure ~\ref{ssh} and in \ref{SN:mSSH}. We can describe the  3,1,$w$-chGNR as a chain of singly-occupied states localized at zigzag edge sites~\cite{Golor2014}, with hopping matrix elements $t$ along the edge, and  width-dependent hopping terms $t_{i}$, $t_{i}^{'}$, and $t_{i}^{''}$ between states at  opposing edges. For very wide ribbons only the edge hopping term $t$ is relevant, and the ribbon's edges enclose  metallic one dimensional bands, as pictured in Fig.~\ref{fig0}b. 
To simulate the emergence of gapped SPT phases during chGNR narrowing, we fitted their VB and CB obtained from 3NN TB simulations using the mSSH model (\ref{SN:mSSH}), and obtained that  a gap opens when the three elements representing the hoping between opposing edges  becomes sizable (Fig.~\ref{ssh}b). Initially, interactions between the diagonal neighbours $t_{i}^{'}$ and $t_{i}^{''}$ (i.e. inter-cell hopping between edges) dominate over confronted zigzag elements (intra-cell hopping,  $t_{i}$). This chiral interaction pattern causes a gapped phase with negative sign, defined from the ${\mathbb{Z}_2}$ invariant as $(-1)^{\mathbb{Z}_2}$, and explains the non-trivial band topology of the 3,1,8-chGNR. 
However, the interaction pattern reverses for narrower ribbons, and the  intra-cell hopping element $t_i$ dominates over the others, leading  now to a gapped phase  with positive sign ($\mathbb{Z}_2=0$) (Fig.~\ref{ssh}b), with a SPT phase transition close to the $w=6$ case.

Inspired by the mSSH model, we performed TB simulations for the \textit{z,a,w}-chGNR family. We computed the band structure and the total Zak phase $\gamma_z$ of occupied bands for the set $0 < z \leq 10$, $a < 6$, and $w \leq 12$, comprising  ribbons with chiral angle from 4.5$^{\circ}$ to 80$^{\circ}$ (\ref{SN:STRUCT}). The resulting band gap values $E_g$ and sign $(-1)^{Z_2}$ are represented in  Fig.~\ref{ssh}\textbf{c}.  
The results show that the SPT phase transition found for the 3,1,$w$-chGNRs is a global property of the $z,a,w$-chGNR family. All chiral ribbons show a similar trend with the width: the gap-less phase of wide ribbons,  with edge states as in Fig.~\ref{fig0}a, transforms first into a one-dimensional topological insulating phase and then into a trivial phase below a critical width.

\begin{figure}[!tb]
\centering
\includegraphics[width = 0.9\textwidth]{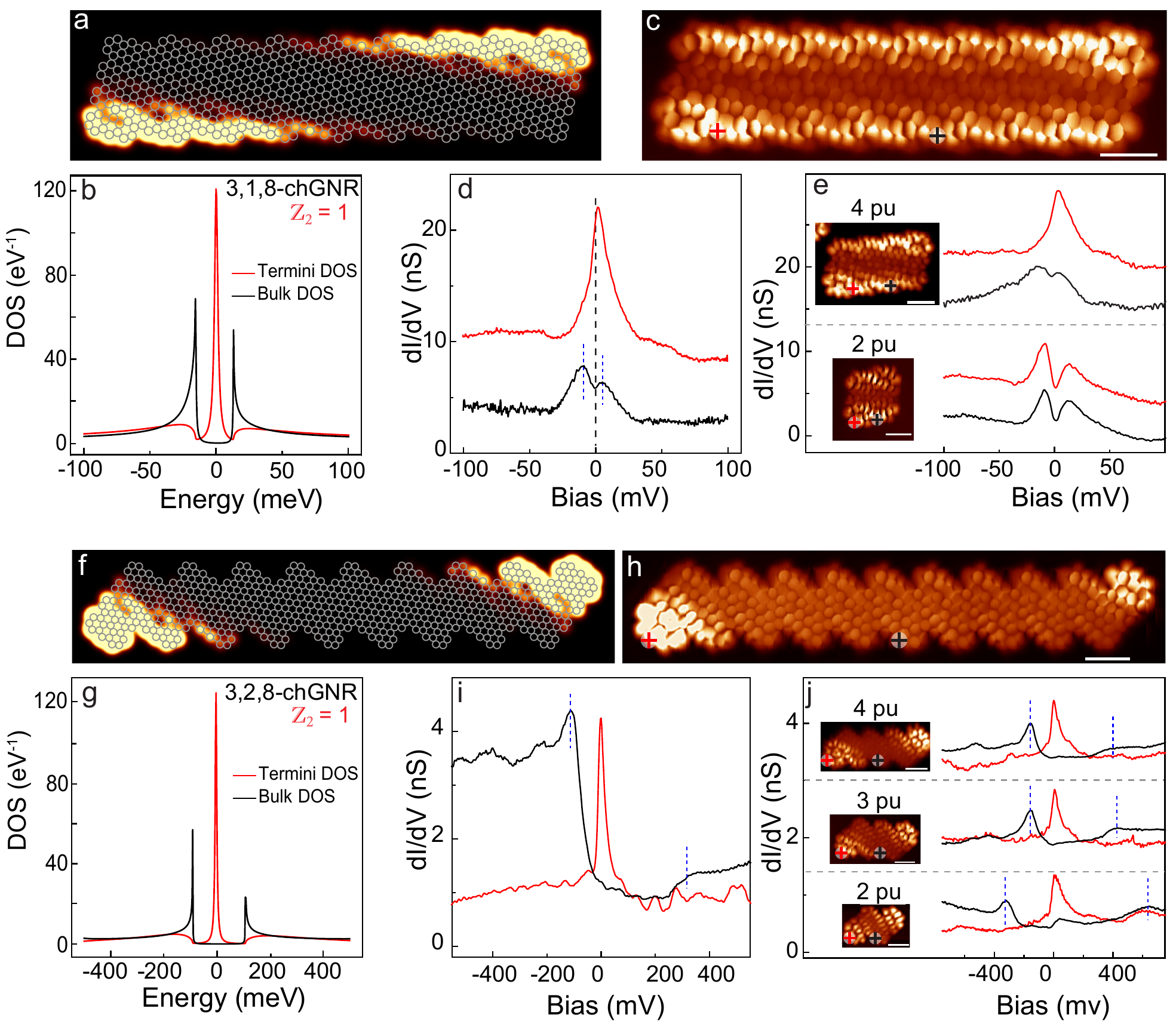}
\caption{\textbf{SPT boundary states of  3,1,8- and 3,2,8-chGNRs:} 
\textbf{a,f}, Zero-energy LDOS distribution of finite 3,1,8-chGNRs and 3,2,8-chGNRs of 9 precursor units obtained with TB simulations and,  
\textbf{b,g}, their corresponding Density of States (DOS) over the center (bulk) and the termini surface showing the localization of zero-energy boundary states. 
\textbf{c,h}, Constant height $dI/dV$ maps ($V=0$ mV) of 3,1,8-chGNRs and 3,2,8-chGNRs with 9 and 10 precursor units respectively and, \textbf{d,i}, $dI/dV$ spectra taken over the termini and edge's center (as indicated with  colored crosses). \textbf{e,j}, Comparison of $dI/dV$ spectra taken on the termini and on center of short ribbons confirm the survival of topological states in ribbons as small as 4 and 2 precursor units for the 3,1,8- and 3,2,8-chGNR, respectively. Insets show  constant height $dI/dV$ maps (V=0 mV). Spectra in \textbf{d,e,j} are shifted for clarity. Blue dashed lines in \textbf{d,e,i,j} indicate the VB and CB onset features. }
\label{TPends}
\end{figure}

\textbf{SPT boundary states:} To experimentally confirm the existence of different gapped SPT phases in this family of chGNRs, we analyze the origin of the persisting low-bias substructure in $dI/dV$ spectra appearing over the zero-bias peaks along the edge (shown in  Fig. \ref{sts}e). It is expected that a non-trivial  bulk-boundary correspondence in topological chGNRs of  finite length leads to  in-gap states localized at the GNR termini and associated to SPT boundary states. Correspondingly, our TB simulations for finite ribbons reproduce  sharp zero-energy states distributed around the ends of a 3,1,8-chGNR (Fig.~\ref{TPends}a,b), which are absent in the narrower ribbons, with  opposite topological class \ref{SN:PTendSlength}. The experimental $dI/dV$ maps measured at low sample bias, like in Fig.~\ref{TPends}c, confirm the presence of these boundary states in 3,1,8-chGNRs,  appearing as a peculiar signal enhancements over the edge's termini, with symmetry and extension similar to the simulated LDOS in Fig.~\ref{TPends}a,b.   
Additionally, $dI/dV$ spectra over these brighter regions show a sharp peak centered at $2$~meV (Fig.~\ref{TPends}d),  and slowly decaying towards the interior of the GNR edge (\ref{318sts}).
In the middle of the ribbon, the VB and CB onsets appear as two peaks at $\sim\pm10$ meV, delimiting a band gap of barely $20\pm 4$ meV. 

%*******************FIGURE 4 *************
%*********

The small band gap $E_g$ of these chGNRs accounts for the slow decay of the end states inside the ribbon (see \ref{SN:PTendSlength}). In short ribbons, end states from opposing termini may overlap and open an hybridization gap $E_{\Delta}$~\cite{Wang2016a} that can hinder the observation of end states when $E_{\Delta}> E_g$. 
In spite of the large spatial extension of the end states in 3,1,8-chGNRs, and the very small gap $E_g$, we found that SPT boundary states survive for  ribbons with only four precursor units (PUs) length, while vanish completely in chGNRs with three PUs or less, whose spectra is fully gapped (Fig.~\ref{TPends}\textbf{e}). 
The survival of end states in short ribbons is confirmed by our TB simulations, which also find the opening of an unusually small hybridization gap (\ref{SN:PTendSlength}). The weak interaction between end states is due to their peculiar distribution in the chiral backbone,  where each end state lays at opposing ribbon's edges (as expected from the mSSH model), thus reducing their overlap in short ribbons.  

Contrasting with the topological character of 3,1,8-chGNR, end states are absent in narrower ribbons,  proving their trivial  insulating phase and confirming the existence of a width controlled topological phase transition in this family of chGNRs. 
For 3,1,6-chGNRs, low bias peaks are spaced by tens of meVs, with a central one pinned  close above the Fermi level (Fig.~\ref{sts}\textbf{e}). As we show in \ref{SN:316sts},  these peaks correspond to valence and conducting bands, discretized in quantum-well states due to the finite length of the ribbon \cite{Carbonell2017,Wang2017,Merino-Diez2018}. 
The peak pinned above  the Fermi level coincides with the VB onset,  partially depopulated   in response to the large electron affinity of the Au(111) substrate \cite{Groning2018,Kimouche2015,Merino2017}. The first peak above the VB corresponds to the CB onset, and no sub-gap features neither signal at the chGNR ends is observed, in agreement with their trivial semiconducting  character.

\textbf{Topological insulating phase of 3,2,8-chGNR:} 
The simulations from Fig.~\ref{ssh}c also illustrate that the size of the chGNR band gaps increase with  the chiral angle, with $E_g$ varying from just a few tens of meV for lower-angle ribbons, to almost one electron-volt for some orientations. This property allows engineering robust topological chGNRs with wider gaps than for the 3,1,8-chGNR. For example, we note that the theoretical gap of the 3,2,8-chGNR shown in Fig.~\ref{ssh} amounts to 199~meV  and is inverted (i.e. $\mathbb{Z}_2=1$). 
Correspondingly, the simulations for finite ribbons of this kind reveal zero-energy topological modes at the termini (Fig.~\ref{TPends}\textbf{f,g}). 

To confirm this topological insulating state, we studied  3,2,8-chGNRs fabricated using the modified precursor \textbf{4}. As depicted in Fig.~\ref{OSS}, the modified halogen substitution of this precursor (at $a=2$ sites)  steers the formation of 3,2,8-chGNRs on a Au(111) surface at elevated temperatures. These ribbons are oriented along a (2,3) vector of the graphene lattice, and alternate three zigzag  with one and a half armchair sites along the edges. The bond resolved STM current image in Fig.\ref{TPends}h confirms the successful OSS of 3,2,8-chGNRs by revealing its characteristic ring structure over the bulk part of the ribbon. However, the STM image also reproduces over the edges a characteristic signal enhancement the resembles the SPT boundary states in the TB LDOS maps of Fig.~\ref{TPends}f. Furthermore, $dI/dV$ spectra measured  over the chGNR ends shows sharp peaks pinned at zero bias, while over the  bulk region of the ribbon a bare  $\sim$ 300 meV band gap is found (Fig.\ref{TPends}i). These end states are then readily identified as SPT boundary states, as predicted by our TB simulations, thus confirming the $\mathbb Z_2=1$ topological class also in this case.  As for the 3,1,8-chGNRs, these end states are pinned slightly above $E_F$, being partially depopulated. However, due to the larger band gap of this ribbon, these end states are more localized at the terminations (\ref{SN:PTendSlength}), and, hence, they are readily detected in even shorter ribbons, with only two precursors units (Fig.~\ref{TPends}j). 

The wider band gap and the larger degree of localization  augment the potential of ribbons with this chirality to become spin-polarized \cite{Lawrence2020,Friedrich2020} in the charge undoped state and in the presence of electron-electron interactions.  
The generalized behaviour described  here for this  family of  chiral GNRs represents a novel route to manufacture graphene ribbons with metallic edge bands  and to transform them into topological states in graphene platforms.  It further shows that endowing graphene with a chiral interaction is an effective method to induce gapped phases. We envision that this method could be extended not only to other chiral geometries in one-dimensional nanoribbons \cite{Jiang2021}, but also to two-dimensional porous graphene networks \cite{Moreno2018}, or Moiré 2D systems, in which combination of flat-bands with chiral symmetries might lead to novel SPT phases.

\begin{acknowledgements}
 We gratefully acknowledge financial support from
the Agencia Estatal de Investigaci\'{o}n (AEI) from projects No  MAT2016-78293,  PID2019-107338RB and FIS2017-83780-P, and from the Maria de Maeztu Units of Excellence Programme
MDM-2016-0618,
the Xunta de Galicia (Centro singular de investigaci\'{o}n de Galicia, accreditation 2016--2019, ED431G/09), the University of the Basque Country (Grant IT1246-19), the Basque Departamento de Educaci\'on through the PhD scholarship no.~PRE\_2019\_2\_0218 (S.S.), and the European Regional Development Fund.
We also acknowledge funding from the European Union (EU) H2020 program,  through the ERC (grant agreement No.~635919) and FET Open project SPRING (grant agreement No.~863098).

Methods and Supplementary Notes and Figures can be downloaded from \url{https://www.springfetopen.eu/publications/}
 
Correspondence and requests for materials should be addressed to:   D.G.O. ({d$\_$g$\_$oteyza@ehu.eus}),  T.F.~(thomas$\_$frederiksen@ehu.eus), D.P.~(diego.pena@usc.es), or  J.I.P. (ji.pascual@nanogune.eu)
\end{acknowledgements}

 \setstretch{1.1}

%\bibliographystyle{biblatex-nature}
%\bibliographystyle{apsrev4-1}  
%\bibliography{Ref-Manuscript}


\end{document}